\documentclass[aps, pra, reprint, twocolumn, floatfix, superscriptaddress, longbibliograph, footnote]{revtex4-2} 
\usepackage{graphicx}
\usepackage{bm, amsmath, amssymb, bbm}
\usepackage[T1]{fontenc}
\usepackage{hyperref}
\usepackage{tabularx}

\newcommand{\ket}[1]{\left|#1\right>}

\makeatletter
\renewcommand*{\@fnsymbol}[1]{\ifcase#1\or i\or ii\or iii\or iv\or v\or vi\or vii\or viii\or ix\or x\else\@ctrerr\fi}
\makeatother

\begin{document}
    \title{Lumped-element broadband SNAIL parametric amplifier with on-chip pump filter for multiplexed readout}
    
    \author{V. R. Joshi*}
    \thanks{* Equal contributions}
    \email{vidul.joshi@yale.edu}
    \altaffiliation{Current address: Microsoft Corporation}
    \affiliation{Department of Applied Physics, Yale University, New Haven, CT 06520, USA}
    
    \author{S. Hazra*}
    \thanks{* Equal contributions}
    \email{sumeru.hazra@yale.edu}
    \affiliation{Department of Applied Physics, Yale University, New Haven, CT 06520, USA}
    
    \author{A. Z. Ding*}
    \thanks{* Equal contributions}
    \email{zhenghao.ding@yale.edu}
    \affiliation{Department of Applied Physics, Yale University, New Haven, CT 06520, USA}
    
    \author{A. Miano*}
    \thanks{* Equal contributions}
    \email{sandro.miano@yale.edu}
    \affiliation{Department of Applied Physics, Yale University, New Haven, CT 06520, USA}
    
    \author{W. Dai}
    \altaffiliation{Current address: Quantum Machines}
    \affiliation{Department of Applied Physics, Yale University, New Haven, CT 06520, USA}  

    \author{G. Umasankar}
    \affiliation{Department of Applied Physics, Yale University, New Haven, CT 06520, USA} 
    
    \author{A. Kottandavida}
    \altaffiliation{Current address: Amazon AWS}
    \affiliation{Department of Applied Physics, Yale University, New Haven, CT 06520, USA}   
    \author{G. Liu}
    \altaffiliation{Current address: Quantum Circuits Inc.}
    \affiliation{Department of Applied Physics, Yale University, New Haven, CT 06520, USA}  
    \author{L. Frunzio}
    \affiliation{Department of Applied Physics, Yale University, New Haven, CT 06520, USA}  
    \author{M. H. Devoret}
    \email{michel.devoret@yale.edu}
    \affiliation{Department of Applied Physics, Yale University, New Haven, CT 06520, USA}  
    \affiliation{Department of Applied Physics, University of California Santa Barbara, CA 93106, USA}  
    \affiliation{Google Quantum AI., Santa Barbara, CA, USA}  
    \date{\today}
    
\begin{abstract}
We present a SNAIL-based parametric amplifier that integrates a lumped-element impedance matching network for increased bandwidth and an on-chip pump-port filter for efficient pump delivery. The amplifier is fabricated using a single-layer optical lithography step, followed by a single-layer electron beam lithography step. We measure a flat 20 dB gain profile with less than 1 dB ripple across a bandwidth of up to 250 MHz on multiple devices, demonstrating robust performance against variations arising from fabrication and packaging. We characterize the amplifier's linearity by analyzing gain compression and intermodulation distortion under simultaneous multi-tone excitation. We show that the intermodulation products remain suppressed by more than 23 dB relative to the signal tones, even at the 1 dB gain compression point. We further validate its utility by performing simultaneous high-fidelity readout of two transmon qubits, achieving state assignment fidelities of $99.51\%$ and $98.55\%$. The combination of compact design, fabrication simplicity, and performance robustness makes this amplifier a practical device for quantum experiments with superconducting circuits.
\end{abstract}

\maketitle

\section{Introduction}

Superconducting parametric amplifiers are key components in quantum information experiments, as they significantly improve the signal-to-noise ratio by amplifying weak microwave signals with near quantum-limited noise performance \cite{Caves1982}. They have enabled a broad range of applications, such as axion dark matter detection \cite{Brubaker2017, Malnou2019, Backes2021, Uchaikin2024, Jiang2023, Bartram2023, DiVora2023}, single-photon and phonon detection \cite{Mohammad2024, Zobrist2019, Ramanathan2024}, and quantum sensing of electronic spin \cite{Elhomsy2023, Wang2023, Vine2025, Bienfait2015, Bienfait2017}. Particularly, in the rapidly developing field of superconducting quantum information processing \cite{Blais2021}, where quantum measurements have become crucial for real-time quantum error correction \cite{Ofek2016, Krinner2022, Campagne-Ibarcq2020, Sivak2023, Google2023, Acharya2025, Putterman2025}, these parametric amplifiers enable single-shot, high-fidelity readout of superconducting qubits \cite{Siddiqi2006, Vijay2009, Clerk2010, Jeffrey2014, Kurilovich2025}. Therefore, they are playing a fundamental role in establishing superconducting qubits as a leading platform for quantum computing \cite{Devoret2013}. These amplifiers come in two flavors: resonant parametric amplifiers \cite{Yurke1989, Yamamoto2008, Bergeal2010, Sivak2019, White2023, Castellanos-Beltran2007, Eichler2014, Winkel2020} and traveling-wave parametric amplifiers (TWPA) \cite{Macklin2015, OBrien2014, Bockstiegel2014, HoEom2012, Esposito2021}. TWPAs are implemented as nonlinear transmission lines comprising thousands of Josephson elements with engineered dispersion \cite{Gaydamachenko2025}. They generally offer higher gain across multi-gigahertz bandwidths, but early versions exhibited significant gain nonuniformity and excess noise \cite{OBrien2014, Macklin2015}. Recent advances have improved both gain flatness \cite{Qiu2023} and noise performance \cite{Chang2025}, bringing TWPAs on par with resonant amplifiers. Nonetheless, their complex design and fabrication remain major barriers to broader deployment, particularly outside the superconducting platforms. Moreover, achieving wideband impedance matching between TWPAs and their environment continues to be technically challenging \cite{Planat2020, Gaydamachenko2025}.
The resonant amplifiers, on the other hand, are typically realized by embedding Josephson elements into superconducting microwave resonators \cite{Yurke1989, Yamamoto2008, Bergeal2010}, which can be fabricated using standard single-layer optical or electron beam lithography. These amplifiers routinely provide amplification with near-quantum-limited added noise over bandwidths of tens of megahertz \cite{Bergeal2010, Frattini2017, Frattini2018}, with center frequencies tunable over the gigahertz range. Their combination of design and fabrication simplicity with near-quantum-limited noise performance has led to widespread adoption in superconducting quantum computing and quantum sensing applications.

For many cutting-edge superconducting circuit applications, however, amplification over a few hundred megahertz with near-quantum-limited noise is more desirable. These applications include but are not limited to multiplexed readout of superconducting qubit arrays \cite{remm_ip3_2023, Schmitt2014, Heinsoo2018}, material characterizations \cite{Wang2015, Ganjam2024}, and fast readout of qubits \cite{Swiadek2023, Spring2024}. This has spurred efforts to extend the bandwidth of resonant amplifiers using impedance matching networks \cite{Roy2015, Mutus2014, Roy2015, Kaufman2024}. Yet, in many such implementations, the measured gain profiles deviate markedly from design expectations due to the sensitivity of impedance matching to fabrication-induced parameter variations.

In this work, we present a broadband SNAIL-based resonant amplifier that combines a simple design and fabrication process with robust performance. We utilize the three-wave mixing properties of Superconducting Nonlinear Asymmetric Inductive eLements (SNAILs) \cite{Frattini2017} for parametric amplification, while incorporating previous results on optimizing SNAIL parametric amplifiers (SPA) designs, such as power handling~\cite{Sivak2019}, pump efficiency \cite{Dai2024}, and impedance engineering \cite{Moskaleva2024}, into a single unified design. Moreover, {we fabricate the devices with a two-layer standard lithography technique,} making our approach accessible to conventional university fabrication facilities. {We experimentally demonstrate that the optimized SPA} provides $20$ dB gain over a $200$ MHz bandwidth ($250$ MHz of $3$ dB bandwidth) with less than $1$ dB ripple with an input saturation power between $-113$ dBm and $-100$ dBm for the entire bandwidth. Furthermore, we incorporate a dedicated pump port {with a band-pass filter that efficiently couples the pump to the SPA, while rejecting signals at the readout frequency. This filter suppresses pump-induced dephasing of the qubits}. To demonstrate the utility of our design, we perform simultaneous readout of two transmon qubits with fidelities exceeding $99.51\%$ and $98.55\%$, respectively, for an integration time of $400$ ns. We provide a detailed analysis of the resulting intermodulation products arising from higher-order nonlinearities of the SPA, and we demonstrate that these products do not limit our multiplexed readout fidelities. 

\section{Design and fabrication}
The design process can be broken down into three parts: the choice of the nonlinear element, the design of the signal port impedance-matching network, and the design of the pump port filter. While these components are necessarily coupled, they can be engineered separately. For the choice of the nonlinear element, we use an array of Superconducting Nonlinear Asymmetric Inductive eLements (SNAILs) \cite{Frattini2017}, which provides the three-wave mixing capability necessary for parametric amplification. Each SNAIL is a flux-tunable inductively shunted Josephson junction, where a small junction is in parallel with three large junctions. To ensure that the SNAILs have a large third-order nonlinearity over a wide range of external magnetic flux, we choose the ratio between the small and large junctions, $\alpha$ to be 0.1. 
In our design, the critical current of the large junction is chosen to be $I_c = 7 \mu A$~\cite{Frattini2017}.
This flux range also includes the Kerr-free point of the SNAIL  {to minimize Stark shift which is known to cause gain compression in SPA~\cite{Frattini2018, Sivak2019}. }
To enhance the amplifier's dynamic range, we employ an array of $20$ nominally identical SNAILs, effectively diluting the higher-order nonlinearities. \cite{Frattini2018, Eichler2014, FrattiniThesis}.

\begin{figure}[t]
    \centering
    \includegraphics[width=0.48\textwidth]{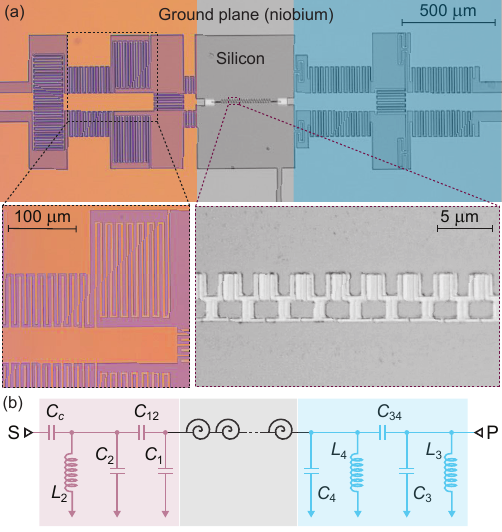}
    \caption{Design of the impedance-matched lumped element SNAIL parametric amplifier with a pump-port filter (a) Micrograph of the device and magnified insets showing  the lumped element interdigitated capacitors and meandering inductors (in orange), the SNAIL array (in gray) and the lumped element pump port filter (in cyan).
    (b) A lumped element circuit model for the amplifier coupled to an impedance-matching network and a pump-port filter.}
    \label{fig:device}
\end{figure}

After choosing the nonlinear element, we focus on the second component, the design of the signal port impedance matching network. This network matches the effective impedance of the pumped SNAIL array to the signal port impedance $Z_0 = 50$ $\Omega$, which determines the gain profile, ripple, and bandwidth of the amplifier. In our design, we aim for a flat gain profile at a center frequency of $\omega_s/2\pi=9$ GHz with $1~\mathrm{dB}$ ripple over a $500$ MHz bandwidth. To do so, we first need to model the effective impedance of the pumped SNAIL array in Fig. \ref{fig:device} (b). {When pumped at twice the signal frequency ($\omega_p \approx 2\omega_s $), the admittance of the array is given by \cite{Sundqvist2014}:}
\begin{equation}{\label{eq:1}}
    Y_{\mathrm{array}}(\omega_s) =  \frac{1}{j \omega_s L_{\mathrm{array}}} - \frac{Z_0 C_1}{L_{\mathrm{array}} |\frac{c_3}{2} \varphi_{p} |^2},
\end{equation}
where $\omega_s$ is the signal frequency, $L_{\mathrm{array}}$ is the linear inductance of the SNAIL array, $\varphi_p$ is the steady-state pump amplitude, and $c_3$ is the coefficient of the third-order non-linearity of the potential energy of the array. {In the presence of a strong pump, the second term in Eq. \ref{eq:1} represents an effective negative effective resistance }
$-|R_{p}| = - {L_{\mathrm{array}} |c_3 \varphi_{p} |^2}/(4Z_0 C_1)$, that produces the gain.
Next, we design a two-pole Chebyshev impedance matching network, where the inductance of the first pole is $L_{\mathrm{array}}$. The target values of the circuit elements can be calculated using standard filter synthesis tables for negative resistance amplifiers \cite{Getsinger1963, Matthaei1980, Naaman2022}, as we discuss in Appendix~\ref{appendix:matching_network}.

The last component is the design of the pump port filter \cite{Dai2024}. To do so, we implement a two-pole filter with a $2~\mathrm{GHz}$ 
passband centered at $20~\mathrm{GHz}$, which provides significant rejection (more than $60$ dB) for tones at signal and idler frequencies while only negligibly attenuating the pump tone. This filter increases the efficiency of pump delivery while preventing unwanted coupling between the pump and signal ports.

The entire circuit is implemented in a coplanar waveguide geometry and fabricated using a single-layer fabrication process with a standard optical lithography step, followed by a single-layer electron beam lithography step. The matching networks and filters are realized using meanders as lumped inductors and interdigitated capacitors with 5 $\mu m$ finger width. The circuit elements, except for the SNAIL array, are fabricated in a single niobium layer on a silicon substrate with a silver backing plane, patterned using standard photolithography. The SNAIL array is subsequently fabricated with electron beam lithography and angled deposition. A detailed fabrication recipe is provided in the Appendix~\ref{appendix:fab}. 

After dicing, the SPAs are packaged in gold-plated aluminum boxes with a superconducting magnet (similar to Ref.~\cite{Sivak2019}) to provide DC flux bias to the SNAIL. The package is magnetically shielded inside a Cryoperm\textsuperscript{\textregistered} box and thermally anchored to the mixing chamber plate of a dilution refrigerator for experimental characterization. The SPA in our main setup serves as the first-stage amplifier in the readout chain of two transmon setups. Each setup consists of a transmon in a 3D aluminum readout cavity. The details of the measurement setup and the device parameters are provided in Appendix~\ref{appendix:setup} and Appendix~\ref{appendix:device_and_readout}, respectively. 

\section{Gain-bandwidth, compression power and measurement efficiency}
\label{sec:gain_bw_p1db}
\begin{figure}[t]
    \centering
    \includegraphics[width=0.48\textwidth]{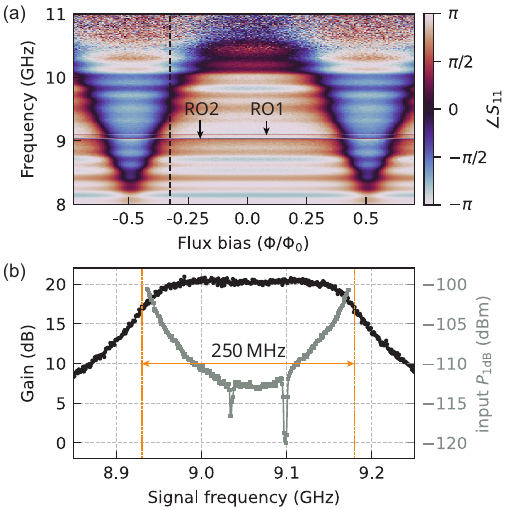}
    \caption{(a) Flux sweep of the phase of the reflection coefficient from the signal. The operating flux point is marked by black vertical dashed line. The response of the two readout resonators $\rm{RO1}$ and $\rm{RO2}$ are also marked. The phase ripples are due to an impedance mismatch in the output line used in this experiment. 
    (b) We show in black, the gain of the amplifier vs frequency. The SPA features a flat-top gain profile with less than $1$ dB ripples and a $3$ dB bandwidth of $250$ MHz. The $1$ dB input compression power within bandwidth of the SPA is shown in gray with respect to a second y-axis on the right. Note that the two sharp dips in the trace are due to the non-linearity of the readout resonators at high power.}
    \label{fig:gain}
\end{figure}

Figure \ref{fig:gain}(a) shows the linear response of the amplifier as a function of the external flux. The two horizontal lines in Figure \ref{fig:gain}(a), labeled `RO1' and `RO2', correspond to the resonant phase response of the two readout cavities. The phase ripples in the color plot are due to impedance mismatch in the output line. As a control experiment, we perform flux sweeps on other devices with identical designs in different cryogenic setups, where the results, included in Appendix \ref{appendix:reproducibility}, do not show such ripples, indicating that the source of these distortions is not from the SPA itself. 

The signal port matching network is designed for the SNAIL
array nominally biased at the dressed Kerr free point \cite{Sivak2019}, $\Phi_{\text{ext}} = 0.35 \Phi_0$, where $\Phi_0$ is the flux quantum.
At this bias point, when pumped at $-50$ dBm through the pump port, the amplifier is expected to produce a power gain of $20$ dB. However, this bias point is subject to change due to experimental nonidealities, such as fabrication defects, junction aging, and the presence of other spurious modes in the package. As a result, we usually perform  {a fine-tuning of the operating conditions by acquiring the gain profile for different combinations of pump frequency, pump power and flux-bias point to obtain a broadband gain with maximum flatness.} Although the amplifier is designed to operate at a single flux bias point for maximum bandwidth, it can still offer an appreciable bandwidth ($> 175$ MHz) for flux-tuning range of $>1$ GHz, ( {see Appendix.~\ref{appendix:tunable_bw} for details.}) allowing multiplexed readout of qubits when the readout resonators are up to hundreds of MHz outside of the maximum gain bandwidth at the optimal flux bias point.

We calibrate the power at the input port of the parametric amplifier  {by measuring the ac Stark shift on the transmons, following} the technique used in \cite{White2023}. Fig.~\ref{fig:gain}(b) shows the power gain of the SPA {(in black)} measured at low input power, $P_{\rm{in}} = -125$~dBm, with a vector network analyzer,  {for the choice of an SPA pump frequency $\omega_p/2\pi = 18.11$ GHz.}  The amplifier achieves a flat top gain profile with less than $1$ dB ripple across a $200$ MHz bandwidth, or 250 MHz when considering the $3$ dB bandwidth of the amplifier. We also characterize the input saturation power, defined by the 1~dB gain compression point $P_{\rm{1dB}}^{\rm{in}}$, as a function of signal frequency across the amplifier bandwidth. 
 {At higher-power, the linear response of the SPA is distorted due to renormalization of the effective inductance of the SNAIL array. 
This renormalization modifies the impedance matching condition and thus deteriorates the flatness of the SPA gain profile. 
The distortion is more prominent at the center of the SPA bandwidth (See Appendix.~\ref{appendix:gain_compression} for details). Therefore, we observe an order of magnitude variation in the compression power across the SPA bandwidth.}
The minimum compression power is found to be $P_{1\rm{dB}}^{\rm{in}}(\omega) = -113$~dBm. At this power, the phase distortion of the amplified signal remains below $\pm 2.5^\circ$ compared to the low-power value. See Appendix \ref{appendix:gain_compression} for more details.
The measured readout efficiencies  {of the two transmons} are $\eta_1 = 0.212$ and $\eta_2 = 0.297$ ($\eta_\mathrm{max} = 0.5$ in phase-preserving operations). From the readout efficiencies and noise visibility ratio~\cite{Frattini2018}, we then estimate the added noise to be $n_{\rm{added}} \lessapprox 0.61$ photons at the SPA input. See Appendix~\ref{appendix:added_noise} for details.

To characterize the pump isolation in our SPA, we measure the Hahn echo coherence times of the two transmons. With the amplifier pump off, we obtain $T_2^E(\mathrm{sys}_1) = 39\pm1~\mu\mathrm{s}$ and $T_2^E(\mathrm{sys}_2) = 76\pm4~\mu\mathrm{s}$. With the pump  {continuously on as a CW tone,} the values are $T_2^E(\mathrm{sys}_1) = 40.\pm1~\mu\mathrm{s}$ and $T_2^E(\mathrm{sys}_2) = 79\pm1~\mu\mathrm{s}$. 
These changes are within the temporal fluctuations of the measured coherence times,  {indicating negligible photon-shot noise arising from the pump leakage. 
We also perform a Ramsey experiment to measure the qubit frequency with and without the CW SPA pump on. We do not observe a noticeable ac Stark shift of the qubit due to the SPA pump. 
These results indicate that we could achieve sufficient pump isolation in our setup without any additional filters or isolators between the readout resonators and the SPA.}

\section{Inter-modulation products}

Gain compression is not the only non-ideality that limits the amplifier performance in the context of multiplexed readout or broadband input signals. When multiple tones are applied to the amplifier simultaneously, the higher-order non-linearity of the amplifier can mix these tones and cause intermodulation distortion (IMD)~\cite{pozar_microwave_2012}.
The {spurious products} interfere with and distort the desired readout signals, compromising the independence of individual readout channels. 
Therefore, characterizing the amplifier’s response to multi-tone inputs is particularly crucial for broadband applications.  
\begin{figure}[t]
    \centering
    \includegraphics[width=0.48\textwidth]{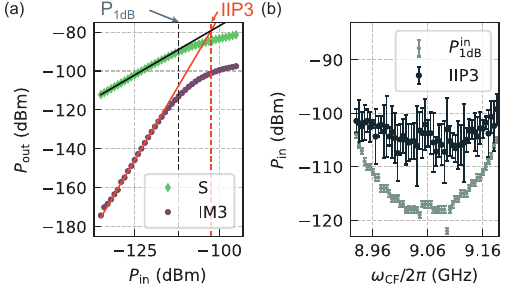}
    \caption{Input third-order intercept points and two tone gain compression. (a) Recorded powers of the output signal tone and a third order mixing product $\omega_{\rm{IM}3} = 2\omega_1-\omega_2$ as a function of the input power of two tones at frequency $\omega_1$ and $\omega_2$. The power of the two input tones are set to be identical, i.e. $P_{\rm{in}}^{s1}=P_{\rm{in}}^{s2}= P_{\rm{in}}^{s}$. Fitted low power asymptotes for the signal (black) and the intermodulation products (orange) and extracted $P^{\rm{in}}_{\rm{1dB}}$ (black, dashed line) and $\rm{IIP}3$ (dashed red line) for a given choice of signal frequencies, centered around $\omega_{\rm{CF}}/2\pi = 8.97$ GHz, and detuned by $0.5$ MHz. (b) Measured two-tone $1$ dB gain compression, $P^{\rm{in}}_{\rm{1dB}}(\omega)$ and $\rm{IIP}3(\omega)$ as a function of the mean frequency $\omega_{\rm{CF}}$ of the two input tones.
    }
    \label{fig:iip3_vs_freq}
\end{figure}

{In three-wave mixing amplifiers, such as SPAs, the leading mixing process annihilates one pump photon at $\omega_p$ and creates two photons, one at the signal frequency $\omega_s$ and another at the idler frequency $\omega_i = \omega_p - \omega_s$. However, the intermodulation products generated by the same three-wave mixing process involving two photons from two different signals, such as $\omega_{\rm{IM2}} = \omega_1\pm\omega_2$ will lie outside of the amplifier band.
This property of the three-wave mixing amplifiers makes the intermodulation products arising from the four-wave mixing process most relevant for amplifier performance.
In the amplifier convention, the order $m$ of a mixing product ${\rm{IM}}m$ is generally denoted by the total number of the signal photons involved in this process~\cite{remm_ip3_2023}. In this section we keep this convention in stating the order of the mixing products. Note that, according to this convention, a four-wave mixing process will generate a mixing product of third order ($\rm{IM}3$).}

Following the approach of Refs.~\cite{Frattini2018, Sivak2019, Kaufman2023}, we characterize the intermodulation distortion by applying two continuous-wave tones detuned by $\Delta/2\pi = 0.5 $ MHz. 
The tones are calibrated to have equal powers at the amplifier input, {$P_{\rm{in}}^{s1} = P_{\rm{in}}^{s2} = P_{\rm{in}}^{s} = P_{\rm{in}}^{\rm{tot}}/2$}. (Note that this definition differs by 3~dB from Ref.~\cite{remm_ip3_2023}).
Inside the amplifier, these tones combine via four-wave mixing to generate an intermodulation product at $\omega_{\rm{IM3}} = 2\omega_1 - \omega_2$, corresponding to the annihilation of two photons at $\omega_1$ and the creation of one photon each at $\omega_2$ and $\omega_{\rm{IM3}}$. We track the output power of this sideband and the amplified signals as functions of input power. We use a signal analyzer with a $1$ Hz resolution bandwidth and $1$ Hz video bandwidth to measure weak signals at low powers.

\begin{figure}
    \centering
    \includegraphics[width=0.48\textwidth]{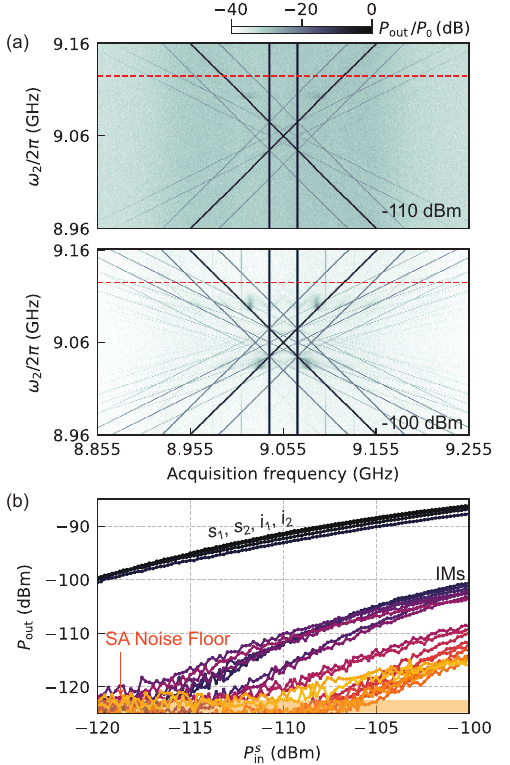}
    \caption{Relevant inter-modulation products within the amplifier bandwidth. (a)
    Measured output spectrum showing signals, idlers, and intermodulation products as a function of the frequency of one input tone $\omega_2$, with the other tone fixed at $\omega_1/2\pi = 9.07$ GHz. Both input tones have equal power, $P_{\rm{in}}^{s1} = P_{\rm{in}}^{s2} = P_{\rm{in}}^s$. The acquired power is normalized to the maximum output power of the desired signals, $P_0 = \max(P_{\rm{out}}^{s1}, P_{\rm{out}}^{s2})$. The output spectra are compared for two input powers, $P_{\rm{in}}^s = -110$ dBm and $P_{\rm{in}}^s = -100$ dBm. The dark spots near the center of the bottom plot are caused by transmon-induced nonlinearity in the readout resonators.
    (b) Input power dependence of the output signals ($s_1$ and $s_2$), idlers ($i_1$ and $i_2$), and $18$ intermodulation products ($\rm{IMs}$), each shown in a different color, along the red line cut in panel (a). The yellow shaded region indicates the signal analyzer’s noise floor for this experiment, for the chosen resolution bandwidth of $50$ kHz.
    }
    \label{fig:imd}
\end{figure}

Fig.~\ref{fig:iip3_vs_freq}(a) shows the power dependence of both the output signal and the intermodulation product when the two tones are centered around $\omega_{\rm{CF}}/2\pi = 8.97$ GHz. Before gain compression, the signal output scales linearly with input power, while the third-order intermodulation product scales cubically. 
To extract the third-order input intercept point ($\mathrm{IIP3}$), we perform linear fits to the low-power regions of the signal and intermodulation outputs. From the intersection of these linear fits, we extract $\mathrm{IIP}3=-102$ dBm (orange dashed line in Fig.~\ref{fig:iip3_vs_freq}(a)). This value is consistent with previous reports on SPAs~\cite{Frattini2018, Sivak2019}. 
When we sweep the tone frequencies across the amplifier bandwidth, we observe that both $P_{1\rm{dB}}^{\rm{in}}$ and $\mathrm{IIP}_3$ are frequency dependent, as illustrated in Fig.~\ref{fig:iip3_vs_freq}(b). However, the variation of $\rm{IIP}3$ is much smaller compared to the variation in compression power; consequently, the ratio ${\rm{IIP}3}/P_{\rm{1dB}}^{\rm{in}}$ varies by an order of magnitude within the amplifier bandwidth. 

While the previous experiment with small detuning is ideal for analyzing pulse distortion, practical multiplexed readout typically involves tone separations ranging from a few MHz to over $100$ MHz. Such a detuning results in intermodulation products that span a broad frequency range. 
To capture all relevant mixing effects, we measure the broadband output spectrum under realistic multiplexing conditions.
We sweep one input tone from $8.96$ to $9.16$~GHz while keeping the other fixed at $\omega_1/2\pi = 9.07$~GHz, covering the amplifier’s $20$~dB flat-top gain region. Both tones are applied at equal powers, varied between $-120$ and $-100$~dBm. The output spectrum is recorded over a $400$~MHz window centered on the amplifier’s bandwidth, using a signal analyzer with a $50$~kHz resolution bandwidth.
At low input powers, only the amplified signals and their corresponding idlers $\omega_i = \omega_p - \omega_s$ are visible. 
As power increases toward the compression point, higher-order intermodulation products appear at frequencies $\omega_{\rm{IM}} = n_p\omega_p + n_1\omega_1 + n_2\omega_2$, with $n_i\in{\mathbb{Z}}$.
Fig.~\ref{fig:imd}(a) shows the measured spectra for $P_{\rm{in}}^s = -110$ and $-100$~dBm. 
We identify the wave-mixing processes for each generated product by analyzing their slopes and  {intercepts (See Appendix~\ref{appendix:labelled_imd} for details)} 
For instance, the output signal $\omega_2$ and its idler $(\omega_p - \omega_2)$ exhibit slopes of $+1$ and $-1$, respectively. 
Two additional slope-$1$ features, $(\omega_2 - 2\omega_1)$ and $(\omega_p - 2\omega_1 + \omega_2)$, arise from third-order mixing.
At higher powers, we also observe vertical features indicating single-tone sideband generation, such as $(2\omega_p - 3\omega_1)/2\pi = 9.01$~GHz and $(3\omega_1 - \omega_p)/2\pi = 9.10$~GHz.

We then fix the input tones at $\omega_1/2\pi = 9.07$~GHz and $\omega_2/2\pi = 9.12$~GHz (red dashed line in Fig.~\ref{fig:imd}) and from the recorded spectra, compare the output powers of $18$ intermodulation products (of order 3 and 5) and the signal-idler pairs. 
Fig.~\ref{fig:imd}(b) shows the power dependence of these intermodulation tones. Notably, intermodulation products of the same order exhibit different output powers depending on their frequencies. This effect arises from the non-uniform $\mathrm{IIP}_3$ across the amplifier’s bandwidth.
A similar trend is visible in the signal and idler traces, which begin to diverge at higher powers, consistent with the amplifier's non-flat gain profile at high power (see Fig.~\ref{fig:spa_03_saturation} in Appendix.~\ref{appendix:gain_compression}).
Despite these distortions, we find that under typical transmon readout conditions ($P_{\rm{in}} \sim -120$~dBm), all intermodulation products remain at least $23$~dB below the signal power, ensuring negligible impact on readout fidelity.


\section{Multi-tone response: Saturation and Frequency crowding}

\begin{figure}
    \centering
    \includegraphics[width=0.48\textwidth]{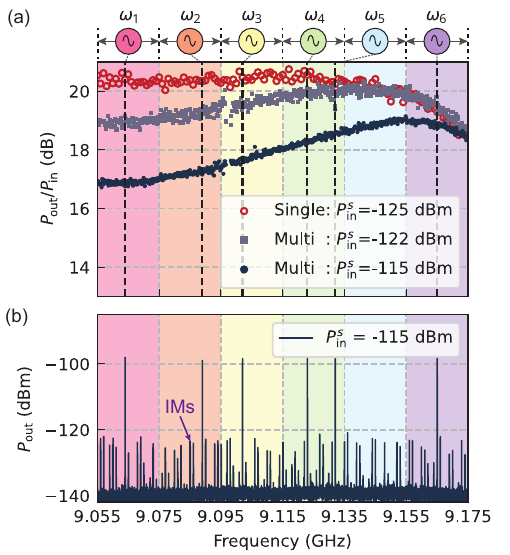}
    \caption{Multi-tone gain response of the broadband amplifier. (a) The amplifier gain is first measured with a single tone at low power ($P_{\rm{in}} = -125$ dBm) across the readout bandwidth (red circles). 
    Then, six CW tones with random initial frequencies within the readout band are applied (black dashed vertical lines). The readout band is divided into six regions, with the local oscillator (LO) frequency swept through each region, borrowing tones from neighboring windows as needed. The gain curve shows expected compression as input power increases.
    (b) The output spectrum at $P_{\rm{in}}^{s} = -115$ dBm shows densely packed intermodulation products. {Despite the increased number of sidebands, their power remains  $23$ dB below the desired signal.}
    }
    \label{fig:multi_tone_gain}
\end{figure}
To further evaluate the gain compression and frequency crowding of the SPA output,  we increase the number of input tones to emulate a realistic frequency-multiplexed readout setup~\cite{White2023}. 
In a frequency multiplexed setup with degenerate parametric amplifiers, both the signals and the idlers share the same physical circuit. 
As a result, only half of the amplifier's bandwidth is  effectively available as the readout band.
One can choose to place the readout frequencies of all the multiplexed members above or below $\omega_p/2$ or alternate them about $\omega_p/2$ to maximize frequency separation among the readout resonators.

In our experiment, we adopt the first configuration, placing the readout band above $\omega_p/2$.
This band is divided into six equal sub-bands (windows), indicated by colored boxes in Fig.~\ref{fig:multi_tone_gain}(a).
We first measure the small-signal gain across the entire readout band using a weak probe tone of $-125$~dBm, shown by the red circles, which serves as a reference.
As multiple readout channels are allocated to the same amplifier chain, avoiding the collisions with the mixing products gets challenging. To represent a scenario in which the readout frequencies are not designed carefully to minimize such collisions, we randomly assign six frequencies (indicated by the black vertical dashed lines) in the readout band. Note that the fourth window (green) contains two tones, while the fifth window (blue) remains empty, emulating practical multiplexing conditions, where all the resonators could not be uniformly allocated in the frequency space.

Each tone is set to a nominally identical input power of $-122$~dBm, consistent with typical transmon readout conditions in our system.

To reproduce the gain response, we sweep each tone across its allocated window, one at a time, while keeping the other five tones fixed at their original frequencies.
In a window containing two tones (green), we sweep the one nearest to the center; if a window is empty (blue), we borrow a tone from a neighboring window (green). 
The resulting gain response is shown by the gray squares in Fig.~\ref{fig:multi_tone_gain}(a). As expected, we observe gain compression relative to the single-tone reference, consistent with total input power $P_{\rm{in}}^{\rm{tot}} = -114.2$~dBm. Increasing the input power of each tone to $-115$~dBm leads to further compression, as shown by the black dots, in agreement with single-tone saturation behavior at equivalent total input powers in Fig.~\ref{fig:spa_03_saturation}.

However, in frequency multiplexing, amplifier's power handling is not the only relevant metric: this multiplexing also introduces spectral complexity due to intermodulation.
As the number of input tones increases, the number of intermodulation products grows combinatorially, resulting in frequency crowding in the amplifier's output spectrum.
To evaluate this effect, we examine the amplifier's output spectrum at $P_{\rm{in}}^s = -115$~dBm, shown in Fig.~\ref{fig:multi_tone_gain}(b). We observe a dense distribution of third- and fifth-order intermodulation products within the amplifier bandwidth, with frequency spacings ranging from $1$ to $4$~MHz. In practice, the readout frequencies are seldom assigned at random; many of the intermodulation products can be sufficiently spaced from the desired signal frequency by carefully designing the readout resonators. This spacing can be further optimized by the choice of pump detuning.
Notably, we observe that the relative power of these spurious tones remains suppressed by more than $23$~dB, even at a relatively high input power. 
These results demonstrate that, despite $\sim 1$ dB output compression and spectral crowding, the amplifier remains well-suited for frequency multiplexing  for typical readout power levels, up to six transmons.

\begin{figure}
    \centering
    \includegraphics[width=0.48\textwidth]{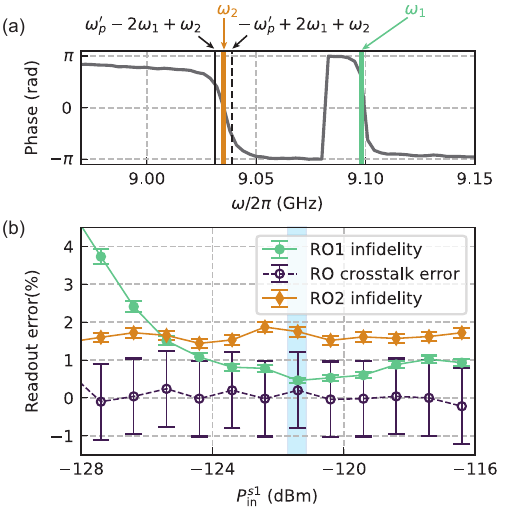}
    \caption{Multiplexed readout performance in presence of frequency collision with a third order mixing product. 
    (a) Phase roll of readout resonators $\rm{RO1}$ and $\rm{RO2}$ relative to intermodulation products $\omega_{\rm{IM3}}^{\pm} = \omega_2 \pm \omega_p \mp 2\omega_1$ under the choice of a SPA pump frequency at $\omega_p'/2\pi = 18.192$ GHz. 
    (b) Readout infidelities and cross-talk error versus the readout power $P_{\rm{in}}^{s1}$, with $\rm{RO2}$ power fixed at $P_{\rm{in}}^{s2} = -119.4$ dBm. Beyond this power, readout-induced leakage dominates the error. 
    The readout infidelity of RO2 remain unaffected as the power of RO1 is increased. Initially, RO1 infidelity decreases as we increase the RO1 power due to improved SNR. As we keep increasing the power, the RO1 infidelity reaches a minimum and then starts to grow due to readout induced leakage. Remarkably, no cross-talk error is observed, even at four times the power corresponding to peak readout fidelity (highlighted by the pale blue box).
    }
    \label{fig:mux_ro_bad_bias}
\end{figure}

\section{Inter-modulation products: Effects on multiplexed readout}
We further demonstrate the amplifier's performance by simultaneously reading out two transmons. We apply a frequency-multiplexed readout pulse with carrier frequencies $\omega_1/2\pi = 9.098$ GHz and $\omega_2/2\pi = 9.035$ GHz, respectively. The ring-down times of the two readout resonators are $\kappa_1^{-1} = 83$ ns and $\kappa_2^{-1} = 24$ ns. The readout duration $\tau_{\rm{RO}}= 400$ ns is set by the response time of the slowest resonator. 
The output signal is demodulated using standard room-temperature electronics (see Appendix~\ref{appendix:setup}).
We do not perform any readout pulse shaping~\cite{hazra_2024} in this experiment to avoid increased spectral bandwidth of the reflected signal originating from the active ramp-up and ramp-down segments of the readout pulse.
We achieve simultaneous single-shot readout fidelities of $99.51\%$ and $98.55\%$ for the two systems, consistent with state-of-the-art transmon readout performance. The lower fidelity of the second system, compared to the first one, is attributed to a suboptimal $\chi/\kappa$ ratio. Further details of the device and readout calibration are provided in Appendix~\ref{appendix:device_and_readout}.

To investigate the impact of frequency collision, we deliberately move the amplifier pump frequency to a suboptimal bias point, $\omega'_p/2\pi = 18.192$ GHz. This pump condition generates images of a pair of third-order intermodulation products $\omega_{\rm{IM3}}^{\pm} = \pm\omega'_p\mp2\omega_1+\omega_2$ within $4$ MHz of the second system's readout frequency. Such a detuning of the sidebands is comparable with the acquisition bandwidth ($\approx\Delta f = 2.5$ MHz) of the readout pulse. The resonator phase roll, the frequency of the readout tone, and the intermodulation products  {for each of the two readouts} are shown in Fig.~\ref{fig:mux_ro_bad_bias}(a). If intermodulation products interfere with the readout signal of system-2 (RO2), they can displace the corresponding readout blobs in the IQ plane. The displacement is set by the amplitude and phase of these intermodulation products, which are themselves determined by the phase of the readout tone for system-1 (RO1).
If the qubit associated with RO1 changes its state, it can thus affect the phase of the intermodulation product and the signal of RO2, leading to a readout crosstalk error~\cite{Heinsoo2018, remm_ip3_2023}:
{
\begin{equation}
    \epsilon_{12} = \Big[P(0_{\rm{RO2}}|g_{\rm{q1}}) + 
    P(1_{\rm{RO2}}|e_{\rm{q1}})-1\Big],
\end{equation}
}
where, $P(m_{\rm{RO2}}|\psi_{\rm{q1}})$ represents the probability of obtaining RO2 outcome as $m_{\rm{RO2}}\in \{0,1\}$, when system-1 is prepared in $\psi_{\rm{q1}}\in\{\ket{g}, \ket{e}\}$.

In this experiment, the intermodulation products are generated by mixing two photons at $\omega_1$ and one photon at $\omega_2$, so the power of the intermodulation product should depend quadratically on the power of RO1. To examine if this frequency collision degrades the multiplexed readout performance in our system, we sweep the power of RO1, while keeping the power of RO2 fixed at $P_{\rm{in}}^{s2} =-119.4$ dBm, beyond which readout-induced leakage in RO2 becomes significant. The resulting readout infidelities (assignment errors) of the two readouts are plotted against the power of the RO1 tone in Fig.~\ref{fig:mux_ro_bad_bias}(b). As expected, initially, RO1 infidelity drops as the signal-to-noise ratio improves at higher power, then it saturates when the readout is no longer limited by SNR, and eventually it increases as the readout-induced leakage dominates the error mechanism. The optimal readout power is shown by the shaded blue region. 
At these powers, the readout fidelities are $99.54\%$ and $98.56\%$, which are comparable to the case without the frequency collisions.
We notice that the readout crosstalk error remains inconsequential for the entire range of readout powers, even when we exceed the optimal readout power by $\sim6$ dB. This implies that the generated intermodulation products are sufficiently suppressed  {for the optimal input powers} used in our readout. 

At present, the maximum readout power is often constrained by unwanted state transitions in the qubit~\cite{sank_2016, khezri_2023, dumas_2024,gusenkova_2021}. However, recent advances in understanding and mitigating these processes,  through increased qubit-resonator detuning to avoid multiphoton resonances, have enabled higher-power readout schemes\cite{Kurilovich2025}. 
At these powers, intermodulation products become more prominent. Additionally, increasing the readout power shortens the readout duration, thereby broadening the acquisition bandwidth. 
As a result, these mixing products interfere strongly with the desired signal, and frequency crowding becomes more significant.
{Nevertheless, for typical parameters used in current state-of-the-art processors~\cite{Acharya2025, kim2023evidence}, our amplifier design provides sufficient power handling ability for multiple readout channels, maintaining adequate isolation between them.}


\section{Discussion and outlook}

In summary, we have demonstrated a broadband SNAIL parametric amplifier that brings together optimized bandwidth, dynamic range, and pump efficiency in two-layer design. 
The resulting device provides $20$~dB gain over a $250$~MHz bandwidth with less than $1$~dB ripple, with input saturation powers between $-113$ and $-100$~dBm across the band. A dedicated pump port filter ensures efficient pump delivery while avoiding internal losses in the signal band.
We further demonstrated the amplifier's performance in a two-qubit readout experiment, achieving high single-shot fidelities of $99.51\%$ and $98.55\%$ in $400$ ns of integration time. We also characterized the amplifier’s behavior under multi-tone excitations and analyzed the intermodulation products, and observed that despite frequency crowding, their suppression was sufficient to preserve readout fidelity and avoid cross-talk.

Looking forward, we can engineer more poles in the impedance matching network for the signal port to increase the bandwidth of the SPAs. Simultaneously, we need to improve the precision of our nanofabrication process to accommodate the reduced margin of error that arises from the increase in pole number. We can also improve our impedance network engineering by performing more detailed characterizations of the impedance mismatches in our devices using techniques such as time domain reflectometry. This will mitigate the discrepancy between the bandwidth of our design and the bandwidth of our actual devices. Moreover, we can also improve the dynamic range of the amplifier by replacing the SNAIL with an RF-SQUID~\cite{Kaufman2023, Kaufman2024}, but at the expense of greater fabrication complexity.

\section*{Author contributions}
\textbf{V.R. Joshi:} Conceptualization (lead); Methodology (lead), Investigation (equal), Data curation (supporting), Formal analysis (supporting), Resources (equal), Validation (equal), Writing-original draft (supporting).
\textbf{S. Hazra:} Methodology (lead), Investigation (equal), Data curation (lead), Formal analysis (lead), Validation (equal), Visualization (lead), Software (equal), Writing-original draft (lead).
\textbf{A.Z. Ding:} Conceptualization (equal), Methodology (lead), Investigation (equal), Formal analysis (supporting), Software (equal), Data curation (supporting), Resources (lead), Validation (equal), Writing-original draft (lead).
\textbf{A. Miano:} Conceptualization (equal), Methodology (equal), Investigation (equal), Data curation (supporting), Formal analysis (supporting), Software (lead), Validation (equal), Writing-original draft (supporting).
\textbf{W. Dai:} Conceptualization (supporting), Resources (supporting), Investigation (supporting).
\textbf{G. Umasankar:} Resources (supporting), Writing-review \& editing (equal).
\textbf{A. Koottandavida:} Software (supporting).
\textbf{G. Liu:} Conceptualization (equal), Methodology (equal), Investigation (supporting), Supervision (equal), Writing-review \& editing (equal).
\textbf{L. Frunzio:} Project administration (equal), Resources (equal), Supervision (equal), Writing-review \& editing (equal).
\textbf{M.H. Devoret} Conceptualization (equal), Investigation (equal), Visualization (supporting), Funding acquisition (lead), Project administration (equal), Supervision (lead), Writing-review \& editing (equal).

\begin{acknowledgments}
This research was sponsored by the Army Research Office (ARO) under grant nos. W911NF-23-1-0051, by the Air Force Office of Scientific Research (AFOSR) under grant FA9550-19-1-0399 and by the U.S. Department of Energy (DoE), Office of Science, National Quantum Information Science Research Centers, Co-design Center for Quantum Advantage (C2QA) under contract number DE-SC0012704. The views and conclusions contained in this document are those of the authors and should not be interpreted as representing the official policies, either expressed or implied, of the ARO, AFOSR, DoE or the US Government. The US Government is authorized to reproduce and distribute reprints for Government purposes notwithstanding any copyright notation herein. Fabrication facilities use was supported by the Yale Institute for Nanoscience and Quantum Engineering (YINQE) and the Yale SEAS Cleanroom.
L.F. is a founder and shareholder of Quantum Circuits Inc. (QCI).
\end{acknowledgments}

\begin{appendix}

\renewcommand{\thefigure}{A\arabic{figure}}
\renewcommand{\thetable}{A\arabic{table}}
\renewcommand{\theequation}{A\arabic{equation}}
\setcounter{figure}{0}  
\setcounter{table}{0}  
\setcounter{equation}{0}  
\section{Matching network synthesis}
\label{appendix:matching_network}
\begin{figure}[b!]
    \centering
    \includegraphics[width=0.48\textwidth]{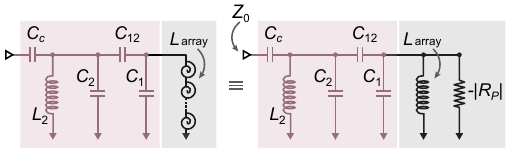}
    \caption{Signal port matching network synthesis using the negative resistance model for driven SNAILs.}
    \label{fig:signal_port_network}
\end{figure}

This section discusses the synthesis of the signal port matching network, which is independent of the pump port. As shown in Fig.~\ref{fig:signal_port_network}, the SNAIL array can be modeled as a negative resistance ($-|R_p|$) in parallel to the linear inductance of the array ($L_{\mathrm{array}}$) in the presence of a strong pump drive. 
We synthesize a filter network to match the port impedance ($Z_0 = 50\Omega$) to $|R_p|$, where the first element of the matching network is $L_\mathrm{arrray}$ \cite{Naaman2019}.

\begin{table}[b!]
    \centering
    \begin{tabular}{|c|c|c|c|c|c|}
        \hline
        $L_{\mathrm{array}}$  & $L_2$& $C_{c}$ &$C_{12}$ & $C_1$ & $C_2$\\
        \hline
        $3.16$ nH & $0.72$  nH & $0.2$ pF &$0.036$   pF & $0.057$  pF  & $0.233$   pF \\
        \hline
    \end{tabular}
    \caption{Values of the components in the matching network.}
    \label{tab:filter_values}
\end{table}
We adopt a lumped element design taking advantage of standard methods of filter network synthesis. We start from the low pass prototype filter coefficients
$g_1 = 0.237197, g_2 = 0.135667, g_3 = 1.119170$~\cite{Getsinger1963}, which correspond to a
Chebyshev profile with 0.1 dB ripple and $20 \mathrm{dB}$ peak gain.  {The maximum gain is determined by the value of $g_3 = Z_0/|R_p|$ and is given by $G_{\mathrm{max}} = |\frac{-|R_p| - Z_0}{-|R_p| + Z_0}|^2$. The relative bandwidth is defined as the ratio of the frequency range over which the gain remains in the range $20 \pm 0.1 \mathrm{dB}$ to the center frequency. For our design, we choose a bandwidth of $w = \Delta f/f = 0.03$, which corresponds to a $3 \mathrm{dB}$ gain bandwidth of $500~\mathrm{MHz}$ at 9 GHz.}

\begin{figure}[t]
    \centering
    \includegraphics[width=0.48\textwidth]{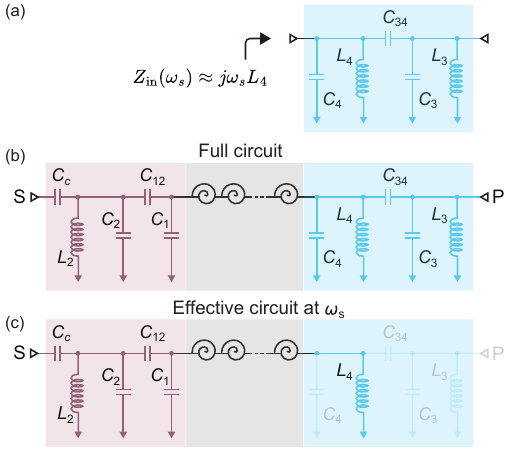}
    \caption{(a) Two-pole band-pass filter design for the pump port. (b) full lumped element schematic of the amplifier. (c) effective amplifier circuit at signal frequencies, $\omega_s$. }
    \label{fig:pump_port_filter_schematic}
\end{figure}

The values of inductors and capacitors in the network are constrained by the geometry. To keep the fabrication simple, we use inter-digitated capacitors with $5 ~\mu m$ finger width and spacing. During the design process, we vary the length and the number of fingers to achieve the target capacitance. The maximum achievable capacitance in this geometry is $0.5~\mathrm{pF}$, beyond which self-inductances of the fingers start to introduce parasitic modes in the frequency band of interest.  Similarly, we implement the inductors as meanders with $5~ \mu m$ trace width and $5~\mu m$ spacing between turns. We vary the number of turns in the design process to achieve the target inductance. The upper bound of the inductance is $1~\mathrm{nH}$ beyond which the effects of parasitic capacitances are no longer negligible.

Taking into consideration the above constraints, we fix the value of the coupling capacitor $C_c = 0.2~\mathrm{pF}$. This capacitor transforms the signal port impedance to $Z_{\mathrm{in}}$, given by
\begin{equation}
    Z_{\mathrm{in}} = Z_0 \frac{1 + q_{\mathrm{in}}^2}{q_{\mathrm{in}}^2} = 200~\Omega,
\end{equation}
where $q_{\mathrm{in}} = \omega_0Z_0C_c$. 

This then fixes the impedance of the second pole,
\begin{equation}
    Z_2 = w\frac{Z_{\mathrm{in}}}{g_3g_2},
\end{equation}
which in turn fixes
the value of the inductor $L_2 = Z_2/\omega_0$. As discussed 
before, since the negative impedance is always embedded in the SNAIL array, the impedance of the first pole is determined by our choice of the SNAIL parameters
$Z_1 = \omega_0 L_{\mathrm{array}}$. The coupling capacitor $C_{12}$ acts as an
impedance inverter whose value is given by
\begin{equation}
    J_{12} = \frac{w}{\sqrt{Z_1 Z_2 g_1 g_2}} = \omega_0 C_{12}. 
\end{equation}
This sets the value of the capacitor $C_{12}$. The remaining
capacitances are then given by 
\begin{align}
    C_1 &= \frac{1}{\omega_0 Z_1} - C_{12} \\ 
    C_2 &= \frac{1}{\omega_0 Z_2} - C_{12} - \frac{C_c}{1+ q_{\mathrm{in}}^2}
\end{align}
The values of the inductors and capacitors are summarized in Table~\ref{tab:filter_values}.
\section{Pump-port filter design}

\begin{figure}[t]
    \centering
    \includegraphics[width=0.48\textwidth]{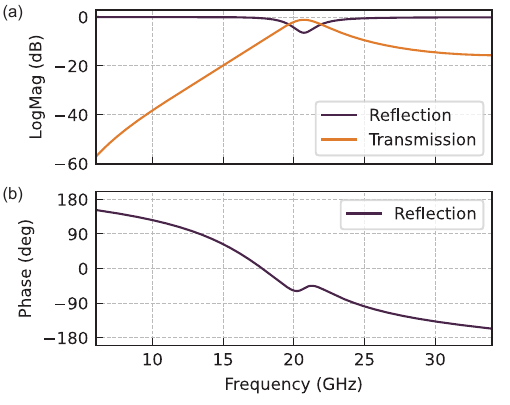}
    \caption{EM Simulation results for pump-port filter. These results show the transmission and reflection amplitude of the pump tone through the pump port filter as a function of frequency in (a) and the phase of the reflected pump tone from the pump port filter as a function of frequency in (b).}
    \label{fig:pump_port_filter_sim}
\end{figure}

The pump-port filter design is optimized to provide a strong coupling at the pump frequency while rejecting the signal frequencies. As a consequence, the filter design also minimizes the pump leakage into the signal port. In the design approach, we also aim to maintain the modularity, so that the matching network and the pump port filter can be independently optimized. 

A simple two-pole bandpass filter, shown in Fig.~\ref{fig:pump_port_filter_schematic}(a) satisfies these requirements. The values of the elements are summarized in Table.~\ref{tab:ppf_values}.
 {To maintain the modularity, we must make sure that, at signal frequency, the input impedance of the pump-port network behaves as a short to the ground. 
This is approximately achieved by placing the shunt resonators near the frequencies in the amplifier gain band.}
This can be seen from the phase response of the reflection coefficient (purple) as shown in Fig.~\ref{fig:pump_port_filter_sim}(b). At the signal frequencies, this phase is close $180^{\circ}$, indicating a short-circuit to ground in the amplifier gain band. In the context of the full amplifier circuit (see Fig.~\ref{fig:pump_port_filter_schematic} (b)), pump port filter shunts the SNAIL
array to ground through the inductor $L_4$ at signal frequencies. The value of this inductance is much smaller than $L_{\mathrm{array}}$. Thus, it only slightly modifies effective $L_{\rm{array}}$ in the signal port design. We account for this renormalization by updating $L_{\mathrm{array}} \rightarrow L_{\mathrm{array}}+ L_{4}$. We iterate this recursive optimization until the circuit parameters converge.

\begin{figure}
    \centering
    \includegraphics[width=0.48\textwidth]{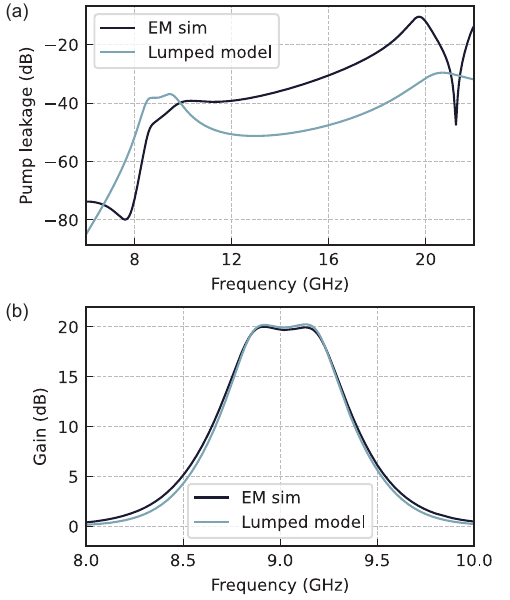}
    \caption{Simulation results for the lumped element SPA. (a) The pump leakage computed from the corresponding simulations. (b)  Simulated gain profile of the simulated SPA from the full EM simulation and with lumped elements model. }
    \label{fig:impedance_matching_sim}
\end{figure}

The orange curve in Fig.~\ref{fig:pump_port_filter_sim}(a) shows the magnitude of transmission through the pump port filter, showing a pass-band close to pump frequencies and $50$-$60~\mathrm{dB}$ rejection at the signal band.  At higher frequencies for the chosen geometry, our lumped element model of our capacitors and inductors breaks down, resulting in different capacitance and inductance across the signal and pump bands. This creates a discrepancy between the pump port filter performance in EM simulations and the lumped element model, as is evident in Figure~\ref{fig:impedance_matching_sim}. This discrepancy is especially prominent for the pump band, as it is much closer to the frequency of the parasitic mode of the capacitor. As a result, for our chosen circuit element geometry, we cannot simultaneously optimize the pump port filter and signal port matching network design. We choose to prioritize signal port engineering in our design because the signal port matching network also rejects tones at pump frequencies. This gives us more tolerance in pump port filter engineering. In our design, we detune the peak of the pass-band of the pump port filter by $2 \mathrm{GHz}$ such that the resulting pump leakage into the signal port remains below $20 \mathrm{dB}$ at the actual pump frequencies.

We model the entire circuit in the AXIEM solver of Cadence AWR, with the SNAIL array treated as a non-linear inductive element.
Fig.~\ref{fig:impedance_matching_sim} (b) shows the simulated leakage across a wide range of frequencies including the pump and signal band.
AWR also has a harmonic balance solver which lets us extract the gain curve of the amplifier for both lumped-element and electromagnetic simulations, as shown in Fig.~\ref{fig:impedance_matching_sim} (a). These simulations only account for the structures on the chip and do not take into account the packaging of the device. Uncontrolled processes like wire-bonding and hand-soldering of the SMA connectors to the printed circuit boards may lead to impedance mismatches. Together with fabrication uncertainties, these can contribute to the discrepancies between the simulated and measured gain bandwidths.

\begin{table}[b]
    \centering
    \begin{tabular}{|c|c|c|c|c|}
        \hline
        $L_{4}$  & $L_{3}$  & $C_{34}$ &  $C_4$  &  $C_3$ \\
        \hline
        ~$0.072$ nH~ & ~$0.072$ nH~ & ~$0.153$ pF~ & ~$0.257$ pF~& ~$0.247$ pF~\\
        \hline
    \end{tabular}
    \caption{Values of the components in the pump port filter.}
    \label{tab:ppf_values}
\end{table}
\section{Device fabrication}
\label{appendix:fab}
The devices were fabricated with a two-layer process on a 300 $\mu$m-thick silicon wafer polished on both sides. First, a 2 $\mu$m-thick layer of silver was evaporated with a Denton Infinity 22 e-beam evaporator at 20 $\mathrm{\mathring{A}}/$s on the back of the wafer to provide all the devices with a ground plane. Prior to the silver evaporation, 20 $n$m of titanium was evaporated at 1 $\mathrm{\mathring{A}}/$s to provide better film adhesion. Afterward, the wafer was sonicated in N-Methylpyrrolidone (NMP), acetone, and isopropyl alcohol (IPA) for 3 minutes each at room temperature. Then, the wafer was blown dry with nitrogen.

Most parts of the microwave circuit in the amplifier, except for the SNAIL array, were fabricated in niobium with photolithography. The niobium layer (150 $n$m) was evaporated on the front side of the silicon wafer using a PVD e-beam evaporator at 10 $\mathrm{\mathring{A}}/$s. Before optical lithography, the substrate surface was primed with HMDS vapor using a TA Series Yield Engineering System (YES) oven for 15 minutes for better resist adhesion and uniformity. The silicon substrate was then coated with SC1805 photoresist using Laurell spin coater WS-400 at 500 rpm for 10 seconds before ramping up to 4000 rpm for 90 seconds for film uniformity. The resist was baked at 115$^{\circ}$C for 1 minute and then loaded into the Heidelberg Maskless Aligner 150 direct laser writer for exposure with a calibrated dose of 70 $\rm{mJ/mm}^2$. Then, the resist was developed with Microposit MF319 developer for 1 minute.

The niobium layer was etched with an Oxford 80 reactive ion etch system. Before the sample was loaded, the chamber was prepped using a cleaning recipe that consisted of two steps: flowing a mixture of O$_2$ and SF$_6$ at flow rates 50 sccm and 10 sccm, respectively, for 10 minutes at 100 W and flowing a mixture of O$_2$ and Ar at 50 sccm and 10 sccm, respectively, for another 10 minutes at 50 W. The niobium etch recipe required flowing SF$_6$ at a rate of 20 sccm and argon at a rate of 10 sccm, generating a pressure of 10 mTorr for 110 seconds to etch through 150 $n$m of metal. After etching, the substrate was cleaned with 3 minutes of sonication in NMP, acetone, and IPA sequentially to remove the photoresist and was blown dry with nitrogen.

The SNAIL loops and the Josephson junctions were patterned using electron-beam lithography. The substrate was spin-coated with 800 nm of MMA (8.5) MAA EL13 at 3000 rpm for 90 seconds and 200 nm of 950K PMMA A4 at 2000 rpm for 90 seconds, with a 1-minute bake at 175$^{\circ}$C after the MMA layer and a 15-minute bake at the same temperature after the PMMA layer. The Dolan-bridge pattern for the Josephson junctions was then written using a Raith EBPG 5200+ at a base dose of 190 $\mu C/\mathrm{cm}^2$. Afterwards, the Ebeam patterns were developed first in 3:1 MIBK:IPA solution for 50 seconds at 25$^{\circ}$C and then in IPA for 10 seconds at room temperature to stop the resist development. The wafer was subsequently blown dry with nitrogen. The Josephson junctions were deposited using a Plassys UMS300 electron-beam evaporator with the double-angled deposition method. After being loaded in the load lock and pumped to base pressure at 2e-7 Torr, the sample was etched with an Ar ion beam at 400 V and 22 mA at $\pm 45^{\circ}$ angles for 34 seconds each. This step cleaned the substrate regions where aluminum was to be deposited and removed the oxide on the niobium sidewalls. Subsequently, the aluminum was evaporated at angles of $\pm 41^{\circ}$ in the evaporation chamber to obtain 35 $n$m and 120 $n$m of aluminum with an oxidation step in the middle using an 85:15 Ar:O$_2$ mixture at 5 Torr for 6 minutes. After the depositions, the surface of the devices was capped \textit{in situ} with a layer of aluminum oxide grown at 10 Torr for 5 minutes before the substrate was unloaded for liftoff.

The liftoff was performed by immersing the substrate in NMP at $75 \ \mathrm{C}$ for 4 hours. Then it was cleaned by sonication in RT NMP, acetone, and IPA for 1 minute each. Then the wafer was blown dry with nitrogen. A layer of SC1827 photoresist was coated on the wafer before it was diced with an ADT ProVectus 7100 dicer. The chips were cleaned by sonication in NMP, acetone, and IPA for 2 minutes each and were blown dry with nitrogen. All the devices were tested by probing at room temperature before cryogenic testing.
\section{Experimental setup for Device A}
\label{appendix:setup}
\begin{figure}[t]
    \centering
    \includegraphics[width=0.48\textwidth]{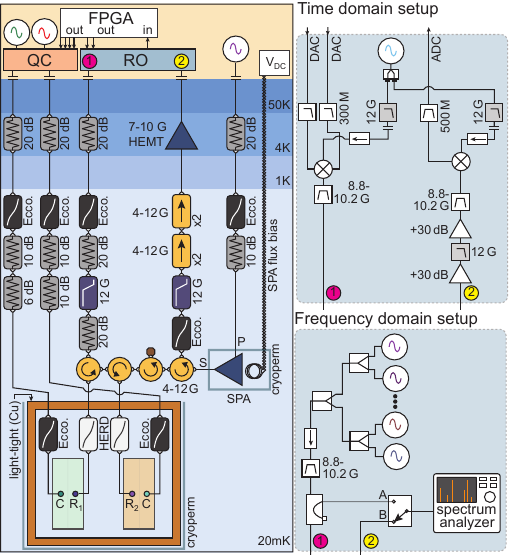}
    \caption{Measurement set up for the multiplexed readout experiment. The boxes labeled ``QC'' and ``RO'' represent the room temperature qubit control and the readout electronics, respectively. On the right, we illustrate the setup for the time-domain (pulsed) measurements (top) and the frequency-domain (CW) measurements (bottom). In labeling the frequency bands of the cryogenic and room temperature components, we use abbreviations `$\rm{G}$' to represent the unit GHz.}
    \label{fig:setup}
\end{figure}

We mount the SNAIL parametric amplifier (SPA) and the two readout cavities at the base temperature ($17$ mK) stage of a dilution refrigerator. Each of the readout cavities houses a fixed-frequency transmon. The two cavities are put in a light-tight copper shield with the inner walls coated with carbon black absorptive layers to eliminate any residual stray photons. The cavities and the quantum-limited amplifier were placed inside a Cryoperm shield to provide protection from stray magnetic fields. (see Fig.~\ref{fig:setup}). Each of the cavities features a separate weakly coupled port for qubit control. 
A double junction circulator is used to multiplex the readout cavities. Another double junction circulator is placed before the SPA to prevent the amplified signal from propagating back to the readout cavities. The SPA output is further amplified by a cryogenic HEMT at the $4$ K stage and two low-noise amplifiers at room temperature. 

For the CW measurements, the inputs for the qubit control lines are terminated with $50$ ohm. The signal and the pump tones are directly generated from microwave CW sources, and the output spectrum is examined with a signal analyzer. We use an FPGA signal processing unit for the time domain measurements that generates IF signals between $0$ MHz and $250$ MHz. We up-convert the signal using IQ modulation and filter spurious sidebands and harmonics before sending it to the readout cavities. Similarly, the amplified output signal is down-converted to an intermediate frequency with respect to the same LO and digitized by the FPGA processing unit at $1$ GS/s. We use a separate CW generator to continuously pump the SPA.

\section{Reproducibility of the matching network}
\label{appendix:reproducibility}

\begin{figure}[tbh]
    \centering
    \includegraphics[width=0.48\textwidth]{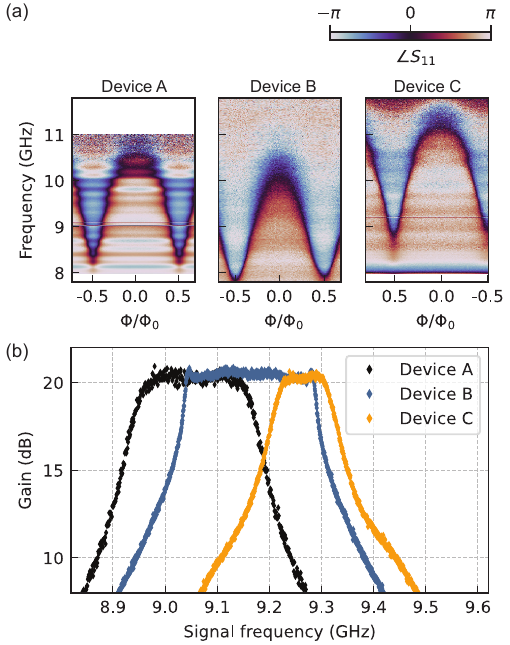}
    \caption{
     {(a) Phase response of the reflection coefficient from the signal port of a three different tested devices, as a function of Flux bias voltage. 
    We show device A, the one presented in the main text, alongside a second device, B, with a flux tuning range closer to the target value and a third device, C, which deviates from the target flux tuning range.
    (b) Corresponding gain profiles of the SPAs, each optimized to produce a flat $20$ dB gain profile with maximum bandwidth. From the observed frequency tuning range, we expect the SNAIL properties of device C to be off from the target value, impacting the impedance matching conditions.} 
    }
    \label{fig:flux_sweep_SPA07}
\end{figure}

To ensure the reproducibility of our results, especially our ability to produce broadband gain up to 100 MHz of bandwidth, we package and cool down three broadband SPAs in different cryogenic setups in the lab. The results are in Fig.~\ref{fig:flux_sweep_SPA07}. Device A is the one presented in the main text, while Device B and C are additional devices we set up in different cryogenic systems for reproducibility testing.

In (a), we show the phase response of the reflection coefficient from the signal port of all three devices as a function of the flux bias. These flux sweeps show slightly different but comparable flux tuning ranges due to junction size variations in the fabrication. In addition, these flux sweeps contain different ripples in the background, owing to the different impedance matching conditions on the output lines of these different cryogenic setups. This distinction indicates that the source of these ripples is not from the SPA itself, but rather part of the external testing conditions. 

We further demonstrate the reproducibility of broadband gain by separately tuning up the three SPAs at their most optimal operating point with maximum gain bandwidth, and we report the optimized broadband gain profiles of these devices in Fig.~\ref{fig:flux_sweep_SPA07} (b). These devices share an identical design for the matching network as well as the pump port filter. However, due to imperfections in the fabrication process, such as suboptimal development of the photoresist and the uneven reactive ion etch across the wafer, the impedance matching network is slightly different from device to device. This variation changes the optimal gain bandwidth, flux operating point, pump frequency, and pump power, resulting in the differences in Figure~\ref{fig:flux_sweep_SPA07} (b). Especially for Device C, at the design frequency, $\omega_s/2\pi = 9$ GHz, the SNAILs are biased near the half flux quantum, thus weakening the three-wave mixing nonlinearity of the device. We are forced to pump it at a frequency different from the target value to produce gain. Despite all of these imperfections, the matching network nonetheless is forgiving enough to produce broadband gain up to a BW >$100$ MHz, showcasing the robustness of our design principles.

\begin{figure}[t]
    \centering
    \includegraphics[width=0.48\textwidth]{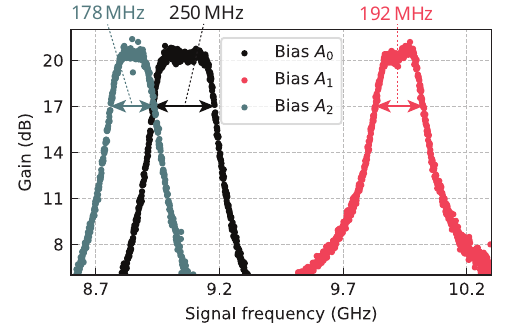}
    \caption{Tunable bandwidth of Device A. The pump frequency is varied over $2$ GHz to tune up flat-top gain profiles, centered around different signal frequencies, between $8.85$ GHz and $9.93$ GHz. The $3$ dB dynamic bandwidth of the SPA for each bias condition is shown above the plot. }
    \label{fig:tunable_bw}
\end{figure}
\section{Tunable bandwidth}
\label{appendix:tunable_bw}

In this section, we provide additional details on the tunable bandwidth of Device A. The characterization focuses on the ability to shift the amplifier’s operational bandwidth across a broad frequency range by varying the pump frequency.
We sweep the pump frequency over a $2$ GHz range and optimized SPA flux bias and the pump power at each step to achieve $20$ dB gain, while maintaining a flat-top gain profile. Across the entire tuning range, between $8.85$ GHz and  $9.93$ GHz, the SPA consistently exhibits a dynamic 3 dB bandwidth greater than $178$ MHz. This robustness highlights the tolerance of the underlying matching network.

\section{Gain compression and phase distortion}
\label{appendix:gain_compression}

\begin{figure}[t]
    \centering
    \includegraphics[width=0.48\textwidth]{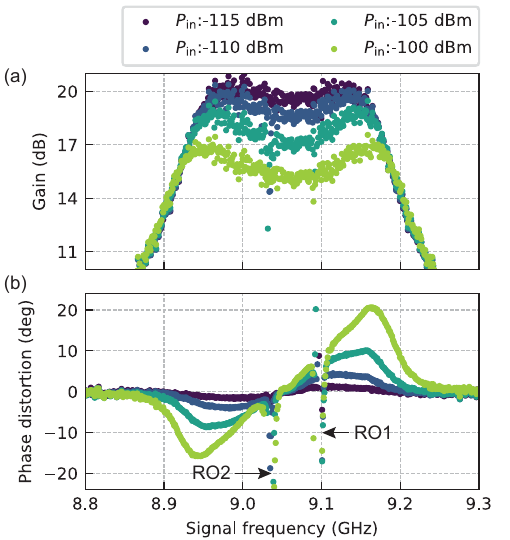}
    \caption{Non uniform (a) gain compression and (b) phase distortion within the bandwidth of the parametric amplifier (Device A) used in the main text.  We measure a higher compression power and a larger phase distortion near the edge of the amplifier bandwidth.      }
    \label{fig:spa_03_saturation}
\end{figure}

To understand the gain compression behavior of our broadband SPA, we measure the gain curve at the optimal operating point of Device A with different input signal powers. As shown in Fig.~\ref{fig:spa_03_saturation} (a), the gain over the flat top is compressed non-uniformly. This can be understood as follows: in general, the gain compression is primarily attributed to the ac Stark shift due to residual Kerr nonlinearity and effective Kerr arising from cascaded processes under drive. The ac Stark shift affects the bare gain and consequently the impedance-matching condition, affecting the flatness of the gain. Consequently, this gives rise to the distorted flat-top gain profile we measure. As we increase the input power, we see an increase in the distortion of the gain curve. This implies that at high power input, signals separated by even small detunings can receive different gains from the SPA, as confirmed by the separation of the signal and idler traces in Fig.~\ref{fig:imd} (b). 

For further characterization of gain compression, we measure the phase distortion across the bandwidth of the same SPA for each input signal power and report our results in Fig.~\ref{fig:spa_03_saturation} (b). The two resonance structures visible in the figure are our two readout cavities. As we increase the input power, we observe increasing distortion in the phase response of the SPA. However, this distortion remains below $\pm 2.5^\circ$ compared to the low-power value for both RO1 and RO2 at the $P_\mathrm{1dB}$ point. This demonstrates the performance of the SPA as a valid amplifier for high-fidelity readout even under high input power.

\section{Measurement efficiency and noise visibility ratio}
\label{appendix:added_noise}
\begin{figure}[tbh]
    \centering
    \includegraphics[width=0.48\textwidth]{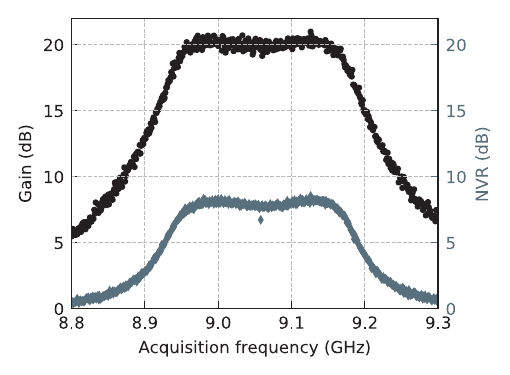}
    \caption{The gain curve and the noise visibility ratio (NVR) of device A is shown on the same plot. We estimate an upper bound on the added noise of the parametric amplifier from the measured efficiency of the output chain and the NVR.
    }
    \label{fig:nvr}
\end{figure}
The measurement efficiency is defined as the ratio between the measurement rate (i.e. the rate at which we obtain information about the qubit state), and the dephasing rate of the qubit under the measurement tone.
The steady state measurement induced dephasing rate for a given power of the measurement tone is expressed as:
\begin{equation*}
    \Gamma_\phi(P) = \frac{\kappa}{2}|\alpha_g(P)-\alpha_e(P)|^2 ,
\end{equation*}
where $\alpha_g$ and $\alpha_e$ are the intracavity pointer states corresponding to the qubit being in the ground and the excited states, respectively.
We compute the dephasing rate with a Ramsey experiment in the presence of a weak measurement tone of varying power. To the leading order (ignoring the self-Kerr of the readout resonator) the dephasing rate is proportional to the power of the measurement tone. We perform a linear fit to extract the power-dependence of the dephasing rate~\cite{Hatridge2013}.

The measurement rate can be measured from the signal-to-noise ratio of the steady-state readout signal as a function of integration time. We fit a Gaussian distribution to each of the resulting $\ket{g}$ and $\ket{e}$ blobs. We define the signal as the square of the distance between the mean of the two distributions, and the noise as the sum of the two variances.
The steady-state measurement rate is given by:
\begin{equation*}
    \mathcal{M}(P) = 2\eta\kappa_{\rm{ext}}|\alpha_g(P)-\alpha_e(P)|^2 
\end{equation*}
Therefore, if the linewidth of the readout resonator is dominated by the coupling to the output line, i.e., $\kappa = \kappa_{\rm{ext}}$, the efficiency is given by $\mathcal{M}(P)/4\Gamma_\phi(P)$~\cite{bultink_2018}.

We measure the readout efficiency of the two systems to be $\eta_1 = 21.2\%$ and $\eta_2 = 29.7\%$ (with $\eta= 50.\%$ being an ideal quantum limited amplifier operating in the  phase preserving mode.) 
This reduction of efficiency is primarily attributed to additional transmission loss (from cables and two more circulator stages) seen by system $1$, compared to system $2$, before the signal reaches the SPA. We calculate this additional transmission loss to be $\delta A = A_1-A_2 = 10\log_{10}(\eta_1/\eta_2) = -1.46$ dB.
Assuming nominally identical cable losses and circulator performances, the total transmission loss between the second cavity and the SPA should be at least $\delta_A/2 + A_{\rm{circ}} = -1.1$ dB. We have assumed the total insertion loss of the double junction circulator, $A_{\rm{circ}} \sim -0.4$ dB, from its cryogenic datasheet.

Accounting for these losses, we scale measurement efficiency  {from the input of the SPA} to be at least $\eta_{\rm{corr}} = \eta_2/10^{(A_2/10)} = 0.38$.
Although this efficiency is primarily determined by the noise added by the SPA, it also includes the noise contribution from the subsequent amplification stages, following the SPA:
\begin{equation}
    \eta_{\rm{corr}} = \frac{N_Q}{N_Q+N_{\rm{SPA}}+N_{\rm{sys}}/G_{\rm{SPA}}},
    \label{eq:efficiency}
\end{equation}
where, $N_Q = 0.5$ corresponds to half a photon of noise. $N_{\rm{sys}}$ and $N_{\rm{SPA}}$ are the system noise power and the added noise of the SPA in terms of the number of photons. $G_{\rm{SPA}}$ is the gain of the SPA.

To put an upper bound on the SPA added noise we measure the noise visibility ratio (NVR), defined as the noise power spectral density with and without the SPA pump on:
\begin{equation}
    \mathrm{NVR} = \frac{N_{\rm{sys}} + G_{\rm{SPA}}\left(N_Q+N_{\rm{SPA}}\right)}{\left(N_Q+N_{\rm{sys}}\right)} 
    \label{eq:nvr}
\end{equation}
We plot the NVR of the SPA across its bandwidth in Fig.~\ref{fig:nvr} along with the SPA Gain curve. 
From Eq.~\ref{eq:efficiency} and Eq.~\ref{eq:nvr}, for $G_{\rm{SPA}} = 20$ dB, we can estimate an upper bound of the SPA added noise to be $0.61$ photons within the bandwidth of the SPA.

\section{IIP3 for input tones with large detuning}
\label{appendix:detuned_iip3}
\begin{figure}[t]
    \centering
    \includegraphics[width=0.48\textwidth]{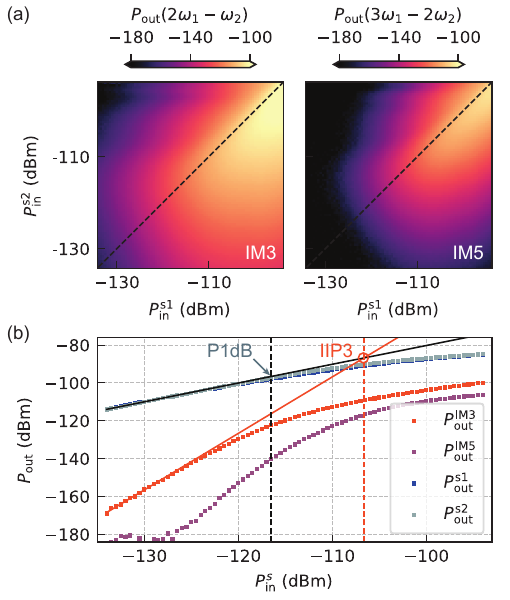}
    \caption{(a) Output power of a third-order and a fifth-order mixing product as a function of the input powers of two signals, $\omega_1/2\pi = 9.07$ GHz and $\omega_2/2\pi = 9.12$ GHz. We track the power of mixing products, at $\omega_{\rm{IM3}} = 2\omega_1-\omega_2$ and $\omega_{\rm{IM5}} = 3\omega_1-2\omega_2$. 
    (b) Line cuts showing the total input power dependence of the mixing products when both the inputs are set to equal powers ($P_{\rm{in}}^{s1} = P_{\rm{in}}^{s2} = P_{\rm{in}}^s$). The mean output power $P_{\rm{out}} = (P_{\rm{out}}^{s1}+P_{\rm{out}}^{s2})/2$ of the two signal tones is linear with respect to the input power at low input powers. Similarly, the third-order mixing products exhibit cubic dependence to input power for small signals. From the power law fits,  the two-tone input $1$ dB compression point $P^{\rm{in}}_{1\rm{dB}}$ and the input third-order intercept point $\rm{IIP}_3$ are extracted to be $-116.5$ dBm and $-106.6$ dBm, respectively, and are shown by the black and the orange solid lines respectively.}
    \label{fig:iip3_iip5}
\end{figure}

In the main text, we compute the $\mathrm{IIP}3(\omega)$ by playing the two input tones with a small detuning of $0.5$ MHz. Here we measure the $\rm{IIP}3$ for two tones separated by a large detuning, similar to Fig.~\ref{fig:imd}.
We apply two tones with frequencies $\omega_1/2\pi = 9.07$ GHz and  $\omega_2/2\pi = 9.12$ GHz at the SPA input and monitor the output power of two in-band third- and fifth-order intermodulation products.
A four-wave mixing process involving two photons at $\omega_1$ and one at $\omega_2$ generates a third-order product at $\omega_{\rm{IM}3} = 2\omega_1-\omega_2$. Similarly, a six-wave mixing process produces a fifth-order product at $\omega_{\rm{IM5}} = 3\omega_1-2\omega_2$. We use a signal analyzer, set to a resolution bandwidth of $1$ Hz, to track the output powers at these frequencies, $\omega_{\rm{IM}3}/2\pi = 9.02$ GHz and $\omega_{\rm{IM}5}/2\pi = 8.97$ GHz.

The measured powers of the two intermodulation products are shown in Fig.~\ref{fig:iip3_iip5}(a), plotted against the input powers of both the tones.
Interestingly, the output power of the sidebands exhibits non-monotonic behavior with respect to $P_{\rm{in}}^{s2}$ at fixed $P_{\rm{in}}^{s1}$. This behavior may arise from power-dependent higher-order corrections to the Kerr nonlinearity~\cite{Sivak2019} or from internal loss mechanisms in the amplifier that are sensitive to input power~\cite{Kaufman2023}.

Next we set the two tones to have the same powers at the input ($P_{\rm{in}}^{s1}= P_{\rm{in}}^{s2} = P_{\rm{in}}^s$) and plot the output powers of the desired signals and the mixing products as a function of the input power $P_{\rm{in}}^s$. The output signal power, the third and the fifth order intermodulation products have slopes $1$, $3$ and $5$ respectively at low input power levels. As described in the main text, we fit these power law dependencies to the measured powers of these output tones and from the intercept points of these lines we extract the input third-order intercept point $\rm{IIP}3 = -106.6$ dBm.  {We could not resolve the fifth-order products above the spectrum analyzer's noise floor at low powers, at which the output power respects the power law.} Thus we could not extract a fifth-order intercept point from our data.

\section{Labeling intermodulation products}
\label{appendix:labelled_imd}
\begin{figure}[tbh]
    \centering
    \includegraphics[width=0.48\textwidth]{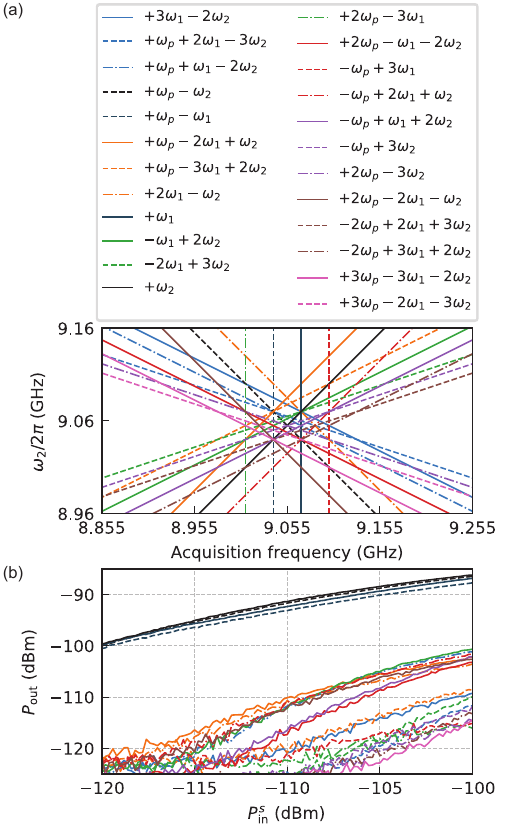}
    \caption{Labeling the signals, idlers and the intermodulation products in the output spectrum. (a) We label all the lower order processes $\omega_{\rm{IM}} = n_p\omega_p + n_1\omega_1 + n_2\omega_2$, with $n_i\in{\mathbb{Z}}$ and $|n_i|\leq3$. (b) We associate these labels to identify specific output products in the experimentally observed spectrum shown in Fig.~\ref{fig:imd}(b)
    }
    \label{fig:imd_labels}
\end{figure}
In Fig.~\ref{fig:imd}(b) of the main text, we have shown the power dependence of the output intermodulation products. In this section, we show how we label the products in terms of the particular process that generates them. In Fig.~\ref{fig:imd_labels} we show all the in-band intermodulation products alongside the signals and the idlers generated by processes like $\omega_{\rm{IM}} = n_p\omega_p + n_1\omega_1 + n_2\omega_2$, with $n_i\in{\mathbb{Z}}$ and $|n_i|\leq3$. We identify these frequencies in the experimentally measured output spectrum and label the products accordingly, while demonstrating their power dependence in Fig.~\ref{fig:imd_labels}(b).

\section{Device properties and multiplexed readout fidelity}
\label{appendix:device_and_readout}

\begin{figure}[t]
    \centering
    \includegraphics[width=0.48\textwidth]{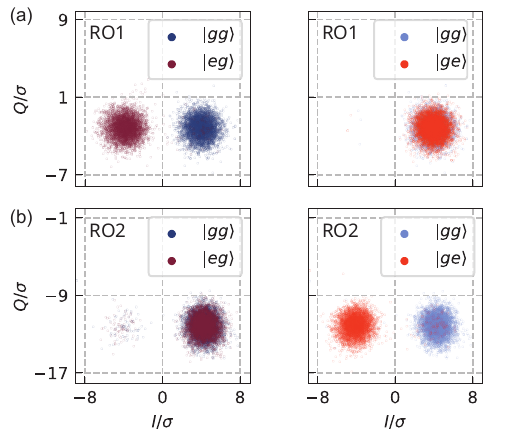}
    \caption{Multiplexed readout performance. Example multiplexed readout signals for the two readouts (a) RO1 and (b) RO2, acquired simultaneously. The state discrimination capability of the two readouts are demonstrated with respect to the prepared states, |gg⟩, |eg⟩, and |ge⟩ of the two qubits. RO1 and
    RO2 identifies the corresponding qubit state with fidelities of $99.51\%$ and $98.55\%$ respectively, while remaining agnostic
    about the state of the other qubit.}
    \label{fig:mux_ro_blobs}
\end{figure}

Our experimental setup consists of two 3D transmons housed in individual rectangular waveguide cavities made of aluminum. The fundamental mode of each cavity serves as the readout resonator of the respective transmon qubit. The \textit{proof-of-principle} readout multiplexing is achieved with an additional cryogenic circulator in the readout chain as shown in Fig.~\ref{fig:setup}.
The device parameters and the coherence times of the two transmons are listed in Table.~\ref{tab:Device parameters}. The linewidth of the resonators and the dispersive shifts are calculated by performing circle fits to the measured resonator responses with the respective transmons prepared in $\ket{g}$ and $\ket{e}$ states.  The coherence values quoted in the table show temporal drift of $\pm10 \mu s$, when measured across several days.

\begin{table}[b]
    \centering
    \begin{tabular}{|l|c|c|}
        \hline
         Properties & System 1 &System 2 \\
         \hline
         Readout frequency, $\tilde\omega_r/2\pi$ (GHz)& 9.0982 & 9.0350\\
         Readout linewidth, $\kappa/2\pi$ (MHz) & 1.9 & 6.6\\
         Qubit frequency, $\omega_q/2\pi$ (GHz)& 4.3050 & 4.4847\\
         Anharmonicity, $\alpha_q/2\pi$ (MHz) & -185 & -184\\
         Total dispersive shift, $\chi_{qr}/2\pi$ (MHz) & -1.8 & -1.3\\
         Qubit relaxation tine, $T_1$ ($\mu$s) & 75 & 82\\
         Qubit Ramsey tine, $T_2^R$ ($\mu$s) & 36 & 68\\
         Qubit Hahn echo tine, $T_2^E$ ($\mu$s) & 40 & 79\\
         \hline
    \end{tabular}
    \caption{Device parameters. Details of the qubit and the readout parameters and the coherence of the two qubits, on which the readout experiments are performed. }
    \label{tab:Device parameters}
\end{table}

We apply a frequency multiplexed $400$~ns long square pulse for the multiplexed readout experiment. 
The amplified and down-converted output signal is acquired for $400$~ns. We apply a fourth order digital Butterworth filter on the two down-converted signals with a cut-off frequency of $10$ MHz and perform digital demodulation. We construct optimal envelopes for the respective signals from the reference trajectories to maximize the discriminability of the qubit states. Finally, the demodulated signals are integrated with the optimal envelopes to obtain a complex number in the phase space.
We acquire $N = 5000$ raw trajectories for each state preparation and readout setting to construct the IQ distribution and compute the readout fidelity, $F_{ii} = 1 -\epsilon_{ii} = [P(0_i|g_i) + P(1_i|e_i)]/2$ with respect to an optimal binary threshold.

Fig.~\ref{fig:mux_ro_blobs} shows the acquired IQ signal of the two readouts, RO1 and RO2, performed simultaneously corresponding to the state preparations, $\ket{gg}$, $\ket{ge}$, and $\ket{eg}$. As shown in Fig.~\ref{fig:mux_ro_blobs}(a), RO1 distinguishes the qubit states, $\ket{gg}$ and $\ket{eg}$ with high fidelity, whereas it remains agnostic about the second qubit's state; consequently, the readout blobs of $\ket{gg}$ and $\ket{ge}$ are completely overlapping. Similar results for RO2 are shown in Fig.~\ref{fig:mux_ro_blobs}(b), highlighting the independence of the two high-fidelity readout channels in our setup.

\end{appendix}

\end{document}